\documentclass[
  ,draft            
  ,mathptm                 
  ]
  {aipproc}

\layoutstyle{8x11double}

\def\lesssim{\mathrel{\hbox{\rlap{\hbox{\lower4pt\hbox{$\sim$}}}\hbox{$<$}}}}
\def\gtrsim{\mathrel{\hbox{\rlap{\hbox{\lower4pt\hbox{$\sim$}}}\hbox{$>$}}}}

\newcommand{\apj}{ApJ}
\newcommand{\apjs}{ApJS}
\newcommand{\apjl}{ApJL}
\newcommand{\aj}{AJ}
\def\mnras   {{\it MNRAS}}

\newcommand{\epcs}{{\rm ergs\,cm^{-2}\,s^{-1}}}

\newcommand{\xte}{{\it RXTE}}	

\newcommand{\chandra}{{\it Chandra}}

\begin{document}

\title{Thermonuclear burst physics with RXTE}

\author{Duncan K. Galloway}{
  address={Massachusetts Institute of Technology}
}

\author{Deepto Chakrabarty}{
  address={Massachusetts Institute of Technology}
}

\author{Andrew Cumming}{
  address={University of California, Santa Cruz}
}

\author{Erik Kuulkers}{
  address={ESTEC}
}

\author{Lars Bildsten}{
  address={University of California, Santa Barbara}
}

\author{Richard Rothschild}{
  address={University of California, San Diego}
}

\begin{abstract}
Recently we have made measurements of thermonuclear burst energetics and
recurrence times which are unprecedented in their precision, largely
thanks to the sensitivity of the {\it Rossi X-ray Timing Explorer}\/ ({\it
RXTE}). In the "Clocked Burster", GS~1826$-$24, hydrogen burns during the
burst via the rapid-proton (rp) process, which has received particular
attention in recent years through theoretical and modelling studies. The
burst energies and the measured variation of alpha (the ratio of
persistent to burst flux) with accretion rate strongly suggests solar
metallicity in the neutron star atmosphere, although this is not
consistent with the corresponding variation of the recurrence time.
Possible explanations include extra heating between the bursts, or a
change in the fraction of the neutron star over which accretion takes
place. I also present results from 4U~1746$-$37, which exhibits regular
burst trains which are interrupted by ``out of phase'' bursts.
\end{abstract}

\maketitle

\section{Introduction}

Unstable thermonuclear ignition of accreted fuel on neutron stars (NSs) in
low-mass X-ray binaries (LMXBs) is triggered once a critical column
density is reached in the fuel layer (e.g. \cite[]{bil00}).  Regular
bursting is surprisingly uncommon, and only one source is known to
consistently burst regularly (GS~1826$-$24; \cite[]{gal03d}).  The
conditions required for regular bursting likely include steady accretion
and uniform spreading of the accreted fuel over the NS surface, as well as
complete fuel consumption.  If the accretion rate is sufficiently steady
the critical density for ignition will be reached after a fixed time. If,
additionally, all the accreted fuel is burnt during each burst, then each
burst will be ignited after the same interval, leading to regular
bursting. Clearly, relaxation of any one of these conditions will lead to
variations in the burst interval, and deviations away from regular
bursting.

Here we present recent results obtained via measurement of thermonuclear
burst properties in GS~1826$-$24 and the globular cluster source
4U~1746$-$37 with \xte. In GS~1826$-$24 we found regular bursting at a
range of accretion rates, which allows us to constrain the composition of
the burning fuel. In 4U~1746$-$37 we found trains of regular bursts
interrupted by bursts which were ``out of phase''. We discuss possible
mechanisms for this phenomenon, as well as future observational tests to
distinguish between them.

\section{Observations and analysis}

We obtained public \xte\/ data from the HEASARC archive at
\url{http://heasarc.gsfc.nasa.gov} and searched for bursts in 1-s binned
Standard-1 data. We found a total of 24 bursts from GS 1826$-$24, and 28
bursts from 4U~1746$-$37. For each burst we extracted full-range
(2--60~keV) time-resolved spectra every 0.25~s from high-time resolution
PCA data modes (GoodXenon or Generic Event, where available), and fitted
an absorbed blackbody to each spectrum after subtracting the persistent
emission. We estimated the observed bolometric flux from the blackbody fit
parameters, and accumulated the measured fluxes to derive the burst
fluence. We also extracted non-burst spectra from both PCA and HEXTE
(15--200~keV) and fitted to absorbed Comptonisation models in order to
estimate the persistent flux level, and hence the accretion rate
$\dot{M}$. Full details of the analysis procedures may be found in
Galloway et al. 2004, in preparation.

We also made simulations to predict ignition conditions for thermonuclear
bursts following \cite[]{cb00}.  We calculate the temperature profile of
the accumulating layer of hydrogen and helium, and adjust its thickness
until a thermal runaway occurs at the base. The temperature is mostly set
by hydrogen burning via the hot CNO cycle, and therefore the CNO mass
fraction $Z$, which we refer to as the metallicity. We assume a constant
level of flux from the crust, at $Q_{\rm crust}=0.1$ MeV per nucleon
\cite[]{brown00}.

\section{Results: GS 1826$-$24}

\begin{figure}
  \includegraphics[height=.5\textheight]{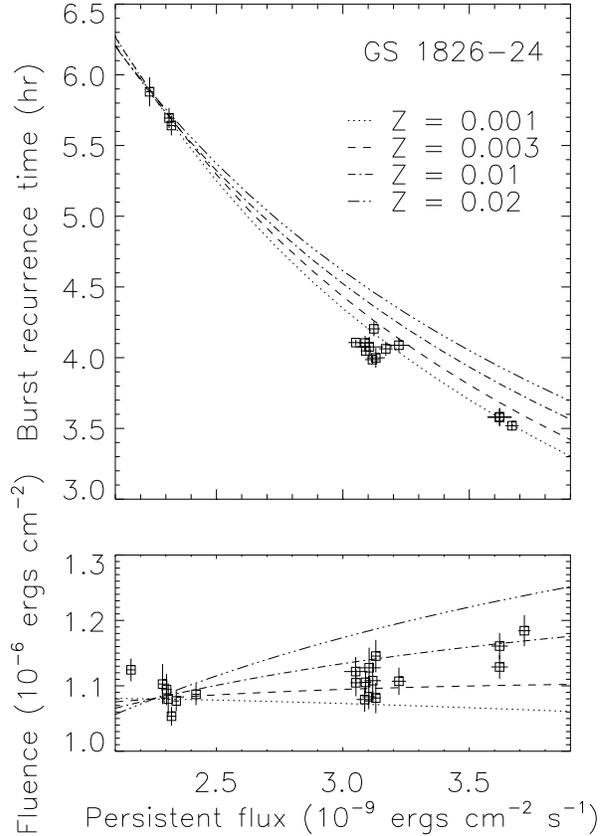}
  \caption{ Variation of the burst recurrence time ({\it upper
panel}) and the burst fluence ({\it lower panel}) as a function of the
estimated bolometric persistent flux in GS~1826$-$24, from \xte\/ measurements
between 1997--2002.
Error bars indicate the $1\sigma$ uncertainties. 
The curves show theoretical calculations for
a range of metallicities: $Z=0.02, 0.01, 0.003,$ and $0.001$. The solid
angle $(R/d)$ and gravitational energy have been chosen in each case to
match the observed fluence and recurrence time at $F_{\rm p}=2.25\times
10^{-9}\ {\rm erg\,cm^{-2}\,s^{-1}}$. For $Z=0.02, 0.01, 0.003,$ and
$0.001$, this gives $R/d=13, 10, 8, 6\ {\rm km}\ @\ 10\ {\rm kpc}$, and
$Q_{\rm grav}=175, 196, 211, 215$ MeV per nucleon, where $Q_{\rm
grav}=GM/R$ is the gravitational energy from accretion.
Reproduced from \cite{gal03d}.
  \label{vstheory} }
\end{figure}

We observed bursts from GS~1826-24 over a factor of 1.7 range in
persistent intensity.  The bursts from a given epoch were consistent with
a single recurrence time, which varied with accretion rate as $\Delta
t\propto\dot{M}^{-1.05}$ (assuming that $\dot{M}$ is proportional to the
persistent intensity $F_{X}$; Fig. \ref{vstheory}).  All the bursts had
similar lightcurves, and exhibited long tails likely powered by rp-process
hydrogen burning.  The burst fluence increased by $\approx5$\% over the
observed range of $F_X$, and the ratio of persistent to burst fluence
$\alpha$ decreased by $\approx 10$\%. The mean value of
$\alpha=41.7\pm1.6$ is in the range expected for mixed H/He burning during
the bursts.

The decrease in $\alpha$ with $\dot M$ implies that stable
burning of hydrogen takes place between the bursts, and suggests solar
metallicity ($Z\approx 0.02$) in the accreted layer. Solar metallicity
models also give good agreement with the observed burst energies, while
low-metallicity models do not.
However, the relatively steep variation in $\Delta t$ with $\dot M$
suggests little variation in the fuel composition at ignition, which in
turn implies low CNO metallicity. Otherwise, hydrogen burning between the
bursts would lead to an increased H-fraction at ignition as $\Delta t$
became shorter, leading to (relatively) delayed ignition and increased burst
fluence.

There are several possible ways to reconcile the solar metallicity
models with the fluence and $\Delta t$ measurements. Firstly, studies of
timing and spectra of LMXBs indicate that $L_X$ is not always a good
indicator of $\dot M$ (e.g.~\cite[]{vdk90}), while we have assumed
$L_X\propto\dot M$ here. Secondly, extra heating of the accumulating layer would
act to reduce the critical mass and bring the observations and theory
into agreement. One possibility is that residual heat from the ashes
of previous bursts heats the layer \cite[]{taam93,woos03},
although time-dependent simulations are required to test this. Thirdly,
if the
fraction of the NS surface covered by fuel changes with
$\dot M$, the changing local accretion rate per unit area could also
reconcile the models and observations. We found that the blackbody
radius $R_{\rm bb}$ in the tail of the bursts decreased by $\approx $20\%
between the observed epochs. If this indicates a change in covering
fraction, it would almost be enough to explain the discrepancy. However,
the covering fraction decrease with $\dot M$ is opposite to the increase
suggested by \cite[]{bil00} to explain trends in burst properties.

\section{4U 1746$-$37}

A well-known dipper and burst source with an orbital period of 5.7~hr
\cite[]{sansom93,homer02}, 4U~1746$-$37 is located in the globular cluster
NGC~6441.  The peak fluxes of the 28 bursts in public \xte\/ observations
were bimodally distributed, with 15 bursts reaching fluxes between
(1.0--$3.5)\times10^{-9}\ \epcs$ (here we refer to these as ``faint''
bursts), and the remaining 13 peaking at between (4.3--$7.1)\times10^{-9}\
\epcs$ (``bright'' bursts).

\begin{figure}
  \includegraphics[height=.28\textheight]{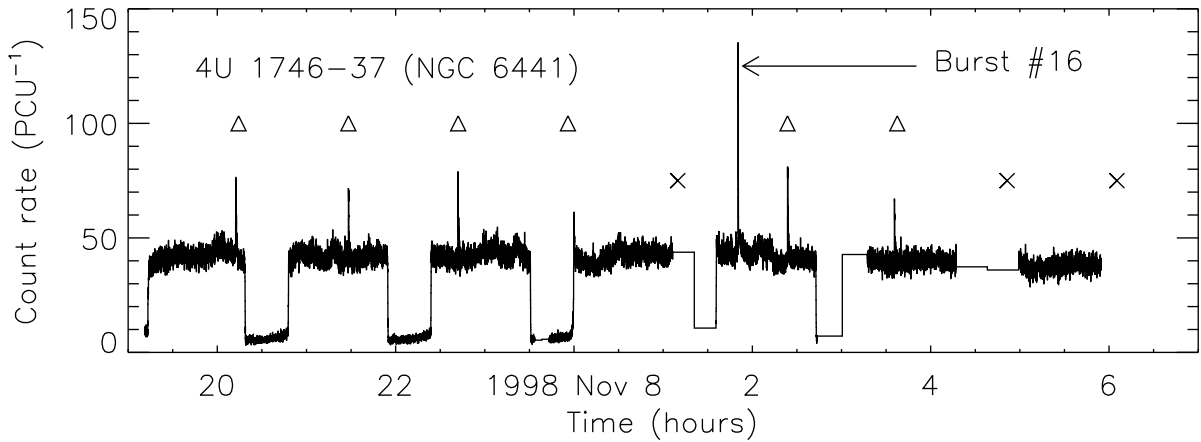}
  \caption{
2--60~keV intensity measured over the field of 4U~1746$-$37 on 1998
November 7--8. The regular dips in the lightcurve are due to occultations
of the source by the Earth due to the $\approx90$~min satellite orbit.
The triangles show the predicted times of bursts according to the
ephemeris determined from bursts \#12--15, 17 and 18. The crosses show
where regular bursts may have occurred, but could not be observed due to
data gaps.  
 \label{example} }
\end{figure}

The bursts appeared to occur regularly on two occasions.  On 1998 November
7th we observed 5 faint bursts (as well as the tail of a 6th burst), each
separated by $\approx1.2$~hr, and with rather uniform properties (fluence,
peak flux, timescale etc.).  However, we also observed a bright burst
which was not consistent with a $\approx1.2$~hr recurrence time. Burst
\#16, which exhibited photospheric radius-expansion (PRE) and reached a
peak of $5.1\times10^{-9}\ \epcs$, occurred 1.84~hr after the previously
observed burst, which was inferred from the observation of a burst tail at
the end of the occultation (\#15; Fig.  \ref{example}).  Neglecting burst
\#16, the remaining bursts were consistent with a rather steady recurrence
time of $1.23\pm0.01$~hr, with an rms error between the observed and
predicted burst times of just 0.034~hr.  The actual recurrence time for
burst \#16 could have been as short as 35~min if a faint burst had
occurred 1.23~hr after burst \#15, although the source was occulted at the
time.

On 1996 October 25--27, when the source was substantially fainter, we also
observed two successive bursts on three instances, again with rather
uniform properties.  However, the majority of the bursts this time were
bright, reaching peak fluxes $\gtrsim 4.5\times10^{-9}\ \epcs$; and
secondly, the recurrence time was 3--3.3~hr. Again, we observed a regular
train of bursts with rather homogeneous properties, interrupted
this time by a faint burst
at a time which was not consistent with the approximately periodic
recurrence of the other bursts.

These deviations from patterns of otherwise regular bursting are puzzling
for two reasons. Firstly, the regular bursting indicates that the critical
column depth for burst ignition occurs on a regular basis; for the
out-of-phase bursts, what causes the fuel layer to ignite prior to
achieving the critical density?  Secondly, assuming constant $\dot{M}$ and
complete consumption of the available fuel in the regular bursts, the
fluence of the out-of-phase bursts appear to be inconsistent with the
amount of fuel accreted since the previous burst.

We suggest three possible explanations for the
observed burst properties:

\noindent {\bf 1. The bright bursts originate from 4U~1746$-$37, but the
faint bursts originate from a second bursting source not positionally
coincident with 4U~1746$-$37}.
Two bursting sources would naturally explain the bimodal distribution
burst properties, as well as the two distinct patterns of regular
bursting.  If the second source was located significantly off the
satellite aimpoint, the observed flux would be reduced due to the decrease
in collimator efficiency with increasing off-axis angle, thus explaining
the relative weakness of the fainter bursts.
However, neither a deep sky map produced from ASM scans of the region
around 1998 November 7--8
nor \chandra\/ observations detected a second source in the
field \cite[]{homer02}, although this could be because the second source
was quiescent at the time.
We also attempted to constrain the origin of the bursts by exploiting
the small differences in pointing between the different proportional
counter units (PCUs) comprising the \xte\/ PCA. We combined the count
rates from individual PCUs from 7 faint bursts observed with similar
spacecraft attitude, in order to increase the signal-to-noise ratio. We
found that the most probable region for the origin of the faint bursts
runs accross the field of view approximately NW to SE, centered on the
position of 4U~1746$-$37.
Thus, we found no evidence that the faint bursts originate from a different
position within the field of NGC~6441.

\noindent {\bf 2. The bright bursts originate from 4U~1746$-$37, and the
faint bursts originate from a second bursting source within NGC~6441}.
Given the low probability of positional coincidence for two unrelated
bursting LMXBs within $\approx 1^\circ$ of each other, combined with the
concentration of NSs in globular clusters, the most likely site
for a second burst source in the field of 4U~1746$-$37\ is NGC~6441. One
final piece of evidence comes from measurements of the asymptotic
blackbody radius for the bursts.  If the second bursting source was not
related to NGC~6441, then we would expect the distance to the source to be
different.  For two NSs emitting X-ray bursts at different
distances, we then expect that the blackbody radii should also be
significantly different. Instead, we find that the blackbody radii are
essentially identical for both classes of bursts.
This suggests the second source has approximately the same distance as
4U~1746$-$37, and hence is most likely within NGC~6441.

\noindent{\bf 3. The bursts all originate from a single bursting source in
NGC~6441, 4U~1746$-$37.}
This hypothesis is somewhat difficult to accept, as discussed above,
because of the bimodal distribution of properties of the non-PRE bursts
from the area, the apparently bimodal distribution of burst recurrence
times, and the interruption of regular trains of one type of burst by
bursts of a second type.

One source could produce both types of bursts if it had undergone a
transition from the regime of unstable hydrogen ignition to that of
unstable He ignition.
In 1996 October for the steady bright
bursts from 4U~1746$-$37 we estimate $\alpha\approx50$, while in 1998
November the $\alpha$-value for the steady faint bursts was much higher, at
$\approx220$.  The former value is typical for bursts which burn mixed
H/He \cite[]{gal03d}, while the latter is more typical for pure He bursts.
Thus, the bright steady bursts may arise from mixed H/He burning triggered
by unstable H ignition at very low accretion rates, while the intermittent
faint bursts burn the resulting He ashes; while in 1998 November, when
$\dot{m}$ was higher, the regular faint bursts are triggered instead by
unstable He burning.  We estimate the accreted column at the time of
ignition for the regular bright bursts assuming cosmic abundances (giving
energy generation of $Q_{\rm nuc}=4.4$~MeV per nucleon) to be
$y\approx2.5\times10^{7}\ {\rm g\,cm^{-2}}$. This is consistent with a H
flash but is at the lower limit of the He ignition curve (e.g.
\cite[]{cb00}).  For the faint bursts, assuming they burn pure helium so
that $Q_{\rm nuc}=1.6$~MeV per nucleon, the column is
$y\approx5.2\times10^7\ {\rm g\,cm^{-2}}$ which is consistent with a He
flash.

Additional observations at different accretion rates are a crucial
test of such a mechanism for giving rise to the bursts observed so far.
If the properties of the two classes of bursts can be shown to vary
independently, this would strengthen the case for two separate sources.
High spatial resolution observations with {\it Chandra}\/ or {\it
XMM-Newton}\/ may then be used to test for two distinct origins for the
two classes of bursts.

A final point which presents additional difficulties for
understanding the properties of this source is the overall frequency of
the bursts.  The broadband persistent flux from the source on 1996 October
and 1998 November was $(0.268\pm0.012)$ and $(1.93\pm0.07)\times10^{-9}\
\epcs$, respectively. For a source distance of 11~kpc, these fluxes
correspond to a spherically-averaged accretion rate of 0.025 and $0.18\,
\dot{m}_{\rm Edd}$ (where we assume $\dot{m}_{\rm Edd}=8.8\times10^4\
{\rm g\,cm^{-2}\,s^{-1}}$). For a NS accreting solar metallicity
material with a hydrogen fraction $X=0.7$ at these rates the model
predicts recurrence times of 44 and 4.1~hr, respectively.  That the
observed recurrence times (3.1 and 1.22~hr) were so short in comparison
may be an indication that the accretion does not completely cover the NS
surface, so that the local accretion rate is much higher.

\section{DISCUSSION}

The precise measurements of burst properties possible with \xte\/ allow
tests of burst theory to unprecedented levels of precision. For both
sources discussed here, the bursting behaviour may be significantly
altered because accretion is not taking place over the entire NS
surface. Measurements of blackbody radii from burst spectra offer at best a
qualitative way to measure the area of accretion, unless the deviations of
the spectra from pure blackbodies can be accounted for and the effective
temperature measured accurately. This can likely only be achieved with
measurements by dedicated spectroscopic instruments like {\it Chandra}\/
or {\it XMM-Newton}.  Such studies, when combined with the extensive
archival observations of bursters accumulated by \xte\/ over its lifetime
are thus an excellent way to improve our understanding of burst physics.
Future observations in previously unexplored ranges of accretion rate will
also further constrain the burst physics.

\bibliographystyle{aipproc}   

\end{document}